# Close-reading of Linked Data: a case study in regards to the quality of online authority files

*Ettore Rizza, Anne Chardonnens, Seth van Hooland*

## Introduction

More and more cultural institutions use Linked Data principles[1] to share and connect their collection metadata. In the archival field, initiatives emerge to exploit data contained in archival descriptions and adapt encoding standards to the semantic web[2]. In this context, online authority files can be used to enrich metadata. However, relying on a decentralized network of knowledge bases such as Wikidata[3], DBpedia[4] or even Viaf[5] has its own difficulties. This paper aims to offer a critical view of these linked authority files by adopting a *close-reading* approach. Through a practical case study, we intend to identify and illustrate the possibilities and limits of RDF[6] triples compared to institutions' less structured metadata.

Our paper is an invitation to travel in an unexpected way: diving through the Linked Open Data cloud[7] by a "thought experiment". Let's suppose that we have a smart robot able to jump from one dataset of RDF triples to another, for example using SPARQL endpoints, and let's call it Alex. Alex is a kind of computerized archivist. We will ask it to use the Linked Data to get information about Henry Carton de Wiart (1869–1951), a famous Belgian personality from the early 20th century. Part of a large noble Walloon family, Carton de Wiart has been minister several times, Prime Minister, president of many councils and organizations, and a lawyer and a writer as well. The Belgian city Liège owes one of its nicknames to one of his books, *La Cité ardente*[8]. Moreover, he has been honored by several awards and was in contact with many well-known personalities, from the French poet Verlaine to the President of America Woodrow Wilson. His life contains therefore enough facets to act as a case study to compare how Linked Open Data can reconstruct a biography compared to more traditional information sources.

---

1  Described in T. BERNERS-LEE, "Design issues: Linked data (2006)", 2011, URL: https://www.w3.org/DesignIssues/LinkedData.html (accessed 14 September 2018). Namely: 1° Use URIs to name (identify) things; 2° Use HTTP URIs so that these things can be looked up; 3° Provide useful information about what a name identifies when it's looked up, using open standards such as RDF, SPARQL, etc; 4° Refer to other things using their HTTP URI-based names when publishing data on the Web.
2  K. F. GRACY, "Archival description and linked data: a preliminary study of opportunities and implementation challenges", *Archival Science*, 15(3), 2015, pp. 239–294.
3  D. VRANDEČIĆ, M. KRÖTZSCH, "Wikidata: a free collaborative knowledge base", *Communications of the ACM*, 57(10), 2014, pp. 78–85.
4  J. LEHMANN, R. ISELE, M. JAKOB, A. JENTZSCH, D. KONTOKOSTAS, P. N. MENDES,… OTHERS, "DBpedia–a large-scale, multilingual knowledge base extracted from Wikipedia", *Semantic Web*, 6(2), 2015, pp. 167–195.
5  M. F. LOESCH, "VIAF (The Virtual International Authority File) – http://viaf.org", *Technical Services Quarterly*, 28(2), 2011, pp. 255–256, https://doi.org/10.1080/07317131.2011.546304
6  Resource Description Framework. D. BRICKLEY, R. V. GUHA, B. MCBRIDE, "RDF vocabulary description language 1.0: RDF Schema. W3C Recommendation", 2004, URL:http://www.W3C.Org/Tr/2004/Rec-Rdf-Schema-20040210.
7  The LOD Cloud Graph, maintained since 2007, is an attempt to map the Linked Open Data ecosystem. The last version (28 August 2018) contains 1224 linked datasets: https://lod-cloud.net/ (accessed 14 September 2018).

Most of the studies about Linked Open Data quality adopt "big data" approaches based on methods such as data mining or network analysis. They tend to analyze the data quality of datasets only through the prism of quantitative and comparative analysis[9]. In this paper, we opt for a close-reading approach focusing on a single individuality, in order to analyze how the triple structure deals with historical/biographical data compared with traditional authority files.

**Continuum**

This paper aims to reflect on the continuum of different documentation practices and methods, from the traditional paper-based narrative to very structured data published as RDF triples in knowledge graphs[10]. At one end are unstructured data, like the national biography of Henry Carton de Wiart, an eight pages well-written text digitized and published online in a PDF version[11]. At the other end stand the most structured data, which corresponds to Linked Data, such as the Wikidata resource for Henry Carton de Wiart[12], its Viaf authority file[13] or its French DBpedia page[14]. In between are established more isolated structured data like RDF triples from archives or libraries repositories[15], archival descriptions in XML[16] or

---

8   H. CARTON DE WIART, *La Cité ardente, avec une préface de M. Henry Bordeaux*, Paris, 1904.
9   A. ZAVERI, A. RULA, A. MAURINO, R. PIETROBON, J. LEHMANN, S. AUER, "Quality assessment for linked data: A survey", in *The Semantic Web*, 7(1), 2012, pp. 63–93.
10  This term, coined and popularized by Google in 2012, refers to knowledge bases structured as RDF graphs.
11  ACADEMIE ROYALE DES SCIENCES, DES LETTRES ET DES BEAUX-ARTS DE BELGIQUE, *Biographie Nationale*, Tome 44, Dernier Supplément, tome XVI, Fasc. 1, Artoisenet-Eracle, 1985, Bruxelles, Retrieved from http://www.academieroyale.be/Academie/documents/FichierPDFBiographieNationaleTome2102.pdf .
12  https://wikidata.org/wiki/Q14990 (accessed 14 September 2018).
13  https://viaf.org/viaf/24623115 (accessed 14 September 2018).
14  http://fr.dbpedia.org/page/Henry_Carton_de_Wiart (accessed 14 September 2018).
15  e.g. Data.bnf.fr: http://data.bnf.fr/12062835/henry_carton_de_wiart/ (accessed 14 September 2018).
16  e.g. the EAC-CPF file from the State Archives of Belgium: https://search.arch.be/eac/xml/eac-BE-A0500_007556_FRE.xml (accessed 14 September 2018).

Wikipedia pages[17].

While the most structured resources display facts (such as a birth date or death date) very clearly, in a machine-readable format, we can wonder if they provide as many details as a more classical format type, such as the national biography. Do they mention his origins from Hainaut, his correspondence with an apostolic vicar living in Brazzaville, or that day in August 1914 when the Belgian king asked him to lead an extraordinary mission to the United States? These questions can be mapped to the central premise of Lev Manovich's book *The Language of New Media*, which stresses the tension between traditional narratives and databases[18]. Through this case study, we aim to extend this thinking and confront various forms of structured sources and to observe to what extent they are consistent and complementary.

**Diving into the LOD cloud**

Alex-the-robot, the main fictional character of our case study, can be seen as a cousin of the modern "robot journalists", these systems able to write short texts—for example weather or financial reports—using structured information[19].

Alex, for its part, would write biographical texts based on information extracted from the Linked Open Data cloud. It will proceed like a historian who collects pieces of information from various sources, to gradually be able to describe a personality. But instead of reading archives or books, Alex would perform queries on the web and then interpret the results, at least if these results are in a machine-readable format such as XML, JSON, any RDF serialization and so on. Even if it does not understand natural language, it is able to write basic sentences, for example that someone was born that year, in that location, and practiced this or that activity.

As a reminder, the principles of Linked Data advocate to use URIs (Uniform Resource Identifiers) as names for things and to use standards such as RDF. For example, the sentence "the artwork *Puppy* held by the Guggenheim Museum was made by Jeff Koons" can be expressed in RDF as follows (using the N-triples notation)[20]:

---

17 e.g. the English version: https://en.wikipedia.org/wiki/Henry_Carton_de_Wiart (accessed 14 September 2018).
18 L. MANOVICH, R. F. MALINA, S. CUBITT, *The Language of new media*, Cambridge, 2001.
19 L. DIERICKX, "News bot for the newsroom: how building data quality indicators can support journalistic projects relying on real-time open data", in *Global Investigative Journalism Conference 2017* Academic Track, Investigative Journalism Education Consortium, 2017.
20 Example borrowed from S. VAN HOOLAND, R. VERBORGH, *Linked Data for Libraries, Archives and Museums: How to clean, link and publish your metadata*, London, 2014.

<http://www.guggenheim.org/new-york/collections/collection-online/artwork/48>

<http://purl.org/dc/terms/creator>

<http://viaf.org/viaf/5035739>.

As we said, Alex's first mission in this experiment is to collect as many triples as possible about Henry Carton de Wiart. Its starting point will be for example DBpedia, often considered as the central point of the Linked Open Data cloud, since many datasets of the LOD cloud are linked to it[21]. Thus, we will ask Alex to go to the Carton de Wiart's English DBpedia page[22], to collect RDF data from this page and then to follow all the links to external databases using, for instance, the owl:sameAs[23] relations.

Figure 1 shows all the links between the knowledge bases used in this experiment. As aforementioned, the entry point is DBpedia. All the paths explored by Alex are visible. From DBpedia, Alex goes to various versions of DBpedia in other languages, then to YAGO, to Freebase (which is discontinued and not used in this experiment) and finally to Wikidata. In our figure, the size of each node is proportional to the number of outgoing links to other databases. The size of Wikidata's bubble means it contains a lot of external links: to VIAF, BNF or the Library of Congress authority ID. Whenever our robot comes in a knowledge base, it collects all the information it finds: Henry Carton de Wiart is a politician; he was born in Brussels; he wrote the novel *La Cité ardente* and so on.

---

21  S. AUER, C. BIZER, G. KOBILAROV, J. LEHMANN, R. CYGANIAK, Z. IVES, "Dbpedia: A nucleus for a web of open data", In *The Semantic Web*, 2007, pp. 722–735.
22  http://dbpedia.org/page/Henri_Carton_de_Wiart (accessed 14 September 2018).
23  https://www.w3.org/TR/owl-ref/#sameAs-def (accessed 14 September 2018).

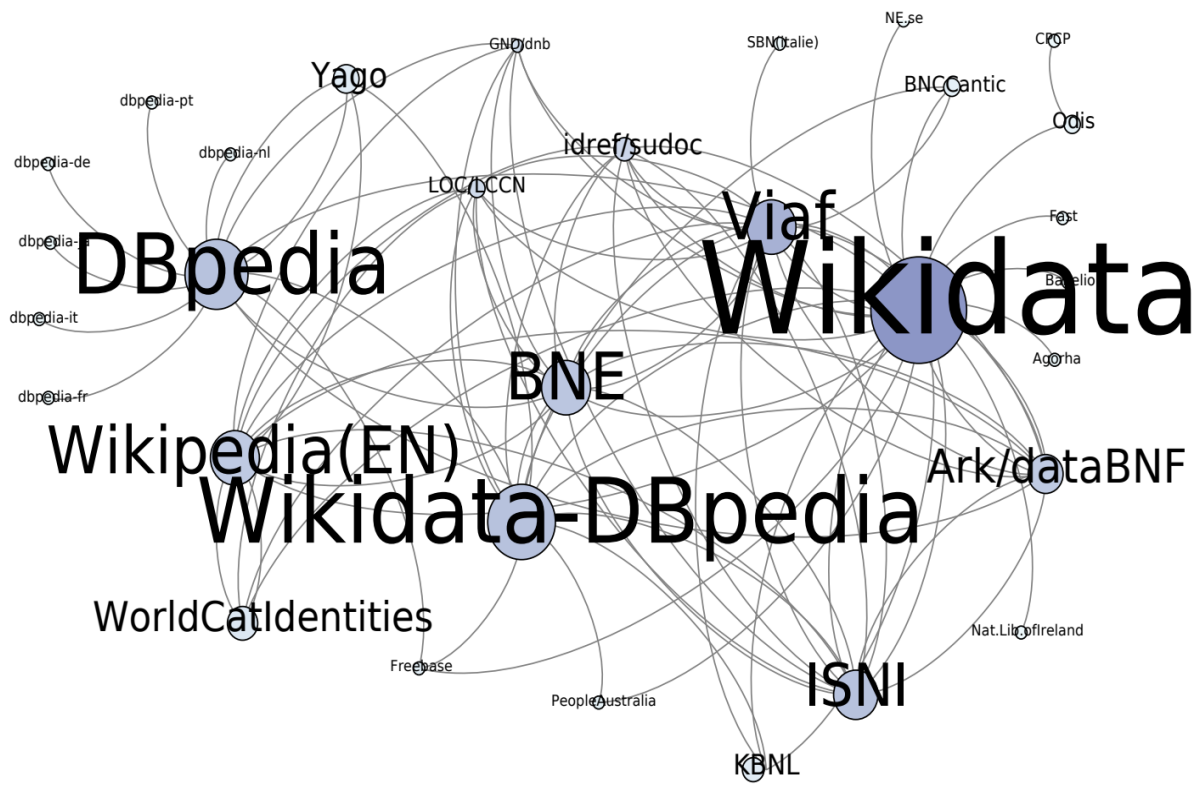

*Figure 1: The part of the Linked Open Data graph traversed during the experiment. The size of each node and label is proportional to the number of outgoing links in the dataset.*

**RDF Triples Harvesting**

During this process, "Alex" (i.e.: the authors) harvested more than 22,000 RDF triples about Carton de Wiart. While this amount can seem huge, it must be noted that the clear majority of these triples are meaningless. RDF can be a very verbose format, requiring sometimes a lot of triples to express something quite basic[24]. Thus, once the useless triples are eliminated, there are about 1,500 triples left, which use some 240 distinct properties.

However, these properties are often redundant, as each database uses their own schemas. To encode a person's date of birth, for example, YAGO uses the property <http://YAGO-knowledge.org/resource/infobox/en/birthdate>, while DBpedia uses

---

24  Especially when institutions such as the Library of Congress provide metadata about their triples, using RDF reification.

<http://dbpedia.org/ontology/birthDate>. In order to facilitate the analysis, we have roughly classified the various properties into eleven empirical classes: affiliations, appellations, category, dates, descriptions, identifiers, locations, professions, relations, works and miscellaneous. Figure 2 shows the proportion of triples in each of these classes.

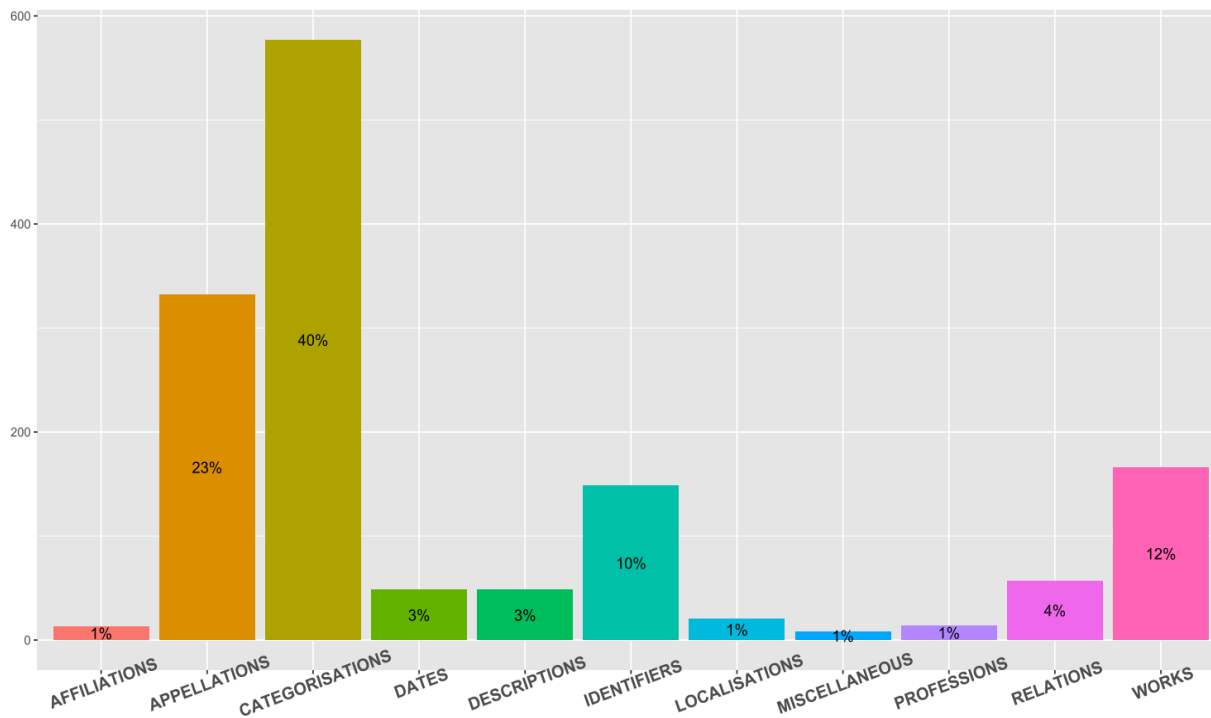

*Figure 2: Distribution of properties manually divided into eleven classes.*

Triples have been placed in their respective category according to the property used. For instance, the property <http://www.w3.org/2000/01/rdf-schematextbackslash#label> corresponds to the "appellation" category, which encompasses all the different labels used to refer to Henry Carton de Wiart. There are many, especially in VIAF, which records the different labels used by libraries: Carton de Wiart, Henry; Carton de Wiart, Henry, Comte, 1869-1951; Carton de Wiart, Henry (1869-1951)… The "category" class is the largest because it includes the (numerous) Wikipedia and YAGO categories —for example <http://YAGO-knowledge.org/resource/wikicat_Walloon_people> or <http://dbpedia.org/resource/Category:Members_of_the_Brussels_Guild_of_Saint_Luke>.

## Semi-Automated Biography

Inspired by the aforementioned reports produced by robot journalists, we tried to see what a biography based on RDF triples would look like. To create this biography, we have synthesized the information contained in the RDF triples collected by Alex. For the sake of clarity, we present in Figure 3 a simplified version, based only on Wikidata and three different versions of DBpedia (EN, FR, NL). This means that this experimental biography does not contain the list of books that Carton de Wiart wrote or collaborated on.

> Henri or Henry Carton de Wiart (first name: Henry or Henri, name: Carton) is a French-speaking Belgian writer and politician, member of the Catholic Party or cdH. He was born on January 31, 1869 in Brussels and died on May 6, 1951 in Uccle. He served as Prime Minister of Belgium from 20 November 1920 to 16 December 1921, replacing Leon Delacroix and replaced by Georges Theunis.
>
> He was a member of the House of Representatives (unknown period) and of the Royal Academy of French Language and Literature of Belgium (from August 9, 1920 to his death, where he was replaced by Hislaire Duesberg). His monarch was Albert the Ist. He belonged to a coalition Catholic Party, PLP, POB. He lived in Rue de la Loi and his residence was "16".
>
> He had as predecessors Albert Lilar, Hendrik Heyman, Henry Jaspar, Leon de Lantsheere, Leon Delacroix and Paul De Groote. He had as successors André_Dequae, Edmond_Rubbens, Emile_Vandervelde, Georges_Theunis, Ludovic_Moyersoen and Paul De Groote.
>
> He had some connection with Liege, Leopold III, Marcel Thiry, the Walloon Movement, Prosper Chicken, the Royal Question, the Wehrmacht, with several legislatures of the House of Representatives, several governments, with Albert Houtart, Gaston Eyskens, Adolf Daens, Jean Duvieusart, Edmond Carton de Wiart, with the Collège Saint-Michel, the moral classes, the circle Leo XIII… He is on the lists of or in the categories of Belgian Prime Ministers, Ministers of Justice, Interior, State, Education in French-speaking Belgium, count, member of the Royal Academy, former deputy, writer of the nineteenth and twentieth century, French writer…

*Figure 3: Here is what an "automated" portrait of Henry Carton de Wiart might look like based on information from the LOD Cloud.*

As shown in Figure 3, all the text fits in less than 20 lines. The result is far from Carton de Wiart's biography from the State Archives of Belgium, let alone the 8-pages biographical note in the National Biography. This confirms our first hypothesis: the overwhelming quantity of triples are often redundant or useless. Most of them express the same elements and often only offer basic information such as birth and death dates. We have highlighted the most problematic excerpts. The text in yellow points out contradictions. Firstly, it is not easy to know how to write Henry. The spelling changes according to the language. In Wikidata, the "given name" property is "Henry", whereas DBpedia uses "Henri". Secondly, the Dutch version of DBpedia[25] wrongly asserts that Carton de Wiart was a member of the "cdH" —a political party created in 2002, after his death. But other knowledge bases correctly indicate that Wiart

---

25  http://nl.dbpedia.org/page/Henri_Carton_de_Wiart (accessed 14 September 2018).

belonged to the Catholic party.

Moreover, the information is not always as structured and clear as we could expect. For example, much of the information in DBpedia or YAGO is not fully explicit. The Dutch version of DBpedia mentions that Carton de Wiart belongs to the Wikipedia's "List of Belgian ministers of Justice". However, this piece of information would not be fully exploitable by a machine, not to mention the fact that the years of his tenure as Minister of Justice are not even indicated. Furthermore, the properties used by YAGO or DBpedia are often poorly described or not described at all. It is therefore difficult to ascertain what the "successor" property (highlighted in green in Figure 3) covers. Successor of whom? For which function?

Similarly, it is doubtful that Carton de Wiart spent his entire life at "16 rue de la Loi", official residence of the Belgian Prime Ministers, presented in DBpedia as his home.

Finally, it appears that this text lacks essential biographical information, for example about his studies or his family. Did he have a wife, children, cousins, parents, siblings? This brings us to the next question: among this missing information, which ones could be easily added and which ones would be more difficult or impossible to translate into RDF?

## Manual "Triplification"

After this first exploration with the help of Alex-the-robot, we wanted to take a close look at the data contained in less structured resources. We tried to do that by acting as if we were an archivist wishing to inject unstructured biographical information into the Linked Data cloud. In other words, we investigate what the options would be to "manually" triplify extra information.

Different types of resources have been used: a biographical note from the *Biographie nationale de Belgique*, an EAC-CPF and an EAD files from the State Archives. During this triplification process, we considered each sentence, one after the other, and tried to extract and translate every single piece of information about Carton de Wiart into RDF triples (subject, predicate, object). For example, a single sentence like "Issu d'une famille de la noblesse catholique, Henry Carton de Wiart effectua ses humanités au collège d'Alost puis au collège Saint-Michel à Bruxelles, avant d'entreprendre des études à l'Université libre de Bruxelles, où il obtint en 1890 son doctorat en droit" need at least seven different triples.

The whole triplification resulted in about 300 statements, containing far more complete and detailed information than the Linked Data triples. During this process, we observed several aspects which can lead to the loss of information, which in turn makes it difficult to deliver a clear and representative description of Henry Carton de Wiart. We have identified three types of challenges, which are respectively related to the data, to the triple structure and to the vocabulary.

- Although the completeness of the data available can constitute a limit, a more challenging point is the granularity: besides practical aspects and available human resources, at what level of detail should a personality and his life be described? For example, in an archival context, should we only use the biographical section of the EAD or also take into consideration other levels of description?

- The triple structure of RDF requires a different way of expressing things. For instance, the translation of one single sentence results sometimes in five or six different triples because of the reification principle.

- If a piece of knowledge can be expressed by words in a biography, like "Henry had four brothers", its full description sometimes has to be inferred from other statements in the RDF language (for example counting the number of triples that use a property like brotherOf).

- Several RDF vocabularies can be used to specify the relationship between persons. Again, we noticed some limits related to granularity. Thus, in the family context, depending on what vocabulary we used, we were able to specify that someone was an uncle, or merely describe the fact that he was a relative. In this case, we would lose details and have uncle and cousin described by the same vague property.

- For social relationship, the vocabulary BIO[26] allows to qualify "acquaintance of", "friend of", "a close friend of", "has met", etc. Although this representation is useful, sometimes it seems quite difficult to evaluate the specific nature of a relationship between two dead people...

- During the triplification process, most of the data has been translated in RDF triples without issues. In a few cases, we lacked vocabulary terms to express more atypical statements, such as details about Henry's personality, or mentions about his activism in the context of social struggles. Obviously, it raises the question of the long tail : to what extent should we create new properties for each very specific case?

**Conclusion and Next Steps**

---

26 https://lov.linkeddata.es/dataset/lov/vocabs/bio (accessed 14 September 2018).

We have seen in this experiment that the Linked Open Data cloud contains a lot of triples about Carton de Wiart, but the total amount of information is quite poor. We have also seen that many other biographical elements can be added as RDF triples, and that there is often a controlled vocabulary that can express them.

But creating our own RDF files requires time and skills. So, another approach might be to automatically feed Wikidata, which can be edited by users unlike DBpedia. In the coming months, we will explore these tracks. We will also try to generalize this first small experiment on a wider panel of personalities and entities, for example organizations or historical events. In other words, we will leave the close reading approach to return to a more classical distant reading.